\documentclass[aps,prl,twocolumn,superscriptaddress]{revtex4}
\usepackage{amsmath}
\usepackage{amssymb}
\usepackage{graphicx}
\usepackage{epsfig}

\newcommand{\om}{\omega}
\newcommand{\Om}{\Omega}

\newcommand{\be}{\begin{equation}}
\newcommand{\ee}{\end{equation}}
\newcommand{\bea}{\begin{eqnarray}}
\newcommand{\eea}{\end{eqnarray}}
\newcommand{\eps}{\epsilon}

\begin{document}

% Use the \preprint command to place your local institutional report
% number in the upper righthand corner of the title page in preprint mode.
% Multiple \preprint commands are allowed.
% Use the 'preprintnumbers' class option to override journal defaults
% to display numbers if necessary
%\preprint{}

%Title of paper
\title{Antimatter interferometry for gravity measurements}

\author{Paul Hamilton}
\author{Andrey Zhmoginov}
\affiliation{Physics Department, University of California, Berkeley, CA 94720, USA}
\author{Francis Robicheaux}
\affiliation{Department of Physics, Auburn University, Auburn, Alabama 36849, USA}
\altaffiliation{Now at Department of Physics, Purdue University, West Lafayette, Indiana 47907, USA}
\author{Joel Fajans}
\altaffiliation{Lawrence Berkeley National Laboratory, One Cyclotron Road, Berkeley, CA 94720}
\author{Jonathan S. Wurtele}
\altaffiliation{Lawrence Berkeley National Laboratory, One Cyclotron Road, Berkeley, CA 94720}
\affiliation{Physics Department, University of California, Berkeley, CA 94720, USA}
\author{Holger M\"uller}
\altaffiliation{Lawrence Berkeley National Laboratory, One Cyclotron Road, Berkeley, CA 94720}
\affiliation{Physics Department, University of California, Berkeley, CA 94720, USA}

%\email[]{Your e-mail address}
%\homepage[]{Your web page}
%\thanks{}
%\altaffiliation{}

%Collaboration name if desired (requires use of superscriptaddress
%option in \documentclass). \noaffiliation is required (may also be
%used with the \author command).
%\collaboration can be followed by \email, \homepage, \thanks as well.
%\collaboration{}
%\noaffiliation

\date{\today}

\begin{abstract}
We describe a light-pulse atom interferometer that is suitable for any species of atom and even for electrons and protons as well as their antiparticles, in particular for testing the Einstein equivalence
principle with antihydrogen. The design obviates the need for resonant lasers through far-off resonant Bragg beam splitters and makes efficient use of scarce atoms by magnetic confinement and atom
recycling. We expect to reach an initial accuracy of better than 1\% for the acceleration of free fall of antihydrogen, which can be improved to the part-per million level.
\end{abstract}

% insert suggested PACS numbers in braces on next line
% insert suggested keywords - APS authors don't need to do this
%\keywords{}

%\maketitle must follow title, authors, abstract, \pacs, and \keywords
\maketitle
\newcommand{\ie}{\emph{i.e.}~}
\newcommand{\eg}{\emph{e.g.}~}

The Einstein Equivalence Principle (EEP) is the basis of gravitational theory. It holds that gravity affects all matter in exact proportion to its mass-energy: All objects experience the same acceleration of free fall $g$, all clocks experience the same gravitational time dilation, and the laws of special relativity hold locally in inertial frames. Matter-antimatter symmetry and the EEP are deep principles of the standard model and gravitational theory, respectively, so that finding them violated would have huge implications. Tests of the EEP have been identified as among the most promising candidates for observable signals of a theory of quantum gravity \cite{Damour2002,KosteleckyTasson}. The EEP is supported by broad experimental evidence for normal, electrically neutral matter. There is compelling experimental and theoretical evidence that antimatter obeys the EEP \cite{indirect}, but these arguments are indirect and are not universally accepted \cite{NoIndirect}; they rely on postulates, e.g., that any gravity anomalies couple to antimatter and matter in a certain way, or that there are no particles and interactions besides those of the standard model and gravity. Since we cannot account for 95\% of the observed gravity in the universe, and since there is much more matter than antimatter in the universe while the accepted laws of physics show matter/antimatter symmetry, we should not presume that the gravitational behavior of antimatter is completely understood. Thus, it is important to explore the gravitational behavior of antimatter in direct experiment.

Neutral antimatter has only recently been trapped in laboratories by the Antihydrogen Laser Physics Apparatus (ALPHA) \cite{Andresen,AndresenNatPhys,AmoleNature} and the Antihydrogen Trap (ATRAP) \cite{Atrap}, while testing the EEP for charged particles \cite{Fairbank} is extremely difficult due to fundamental and practical limitations on how well they can be isolated from the environment.  As a result, the EEP has been directly confirmed neither for antimatter (for which $g=(-0.63\ldots +1.1)$\,km/s$^2$ is compatible with the data \cite{Antig}), nor for charged particles of any kind. The AEGIS collaboration at CERN aims to measure the gravitational acceleration of antihydrogen by a Moir\'e accelerometer, which is now in its final construction phase \cite{Aegis}. A second experiment, GBAR, will drop antihydrogen from a height of 10\,cm and has recently been approved at CERN \cite{Gbar}. Both expect to reach a percent-level accuracy.

Light-pulse matter wave interferometers \cite{AI} have been used to measure, {\em e.g.}, local gravity \cite{Peters}, the gravity gradient \cite{Snadden}, Newton's gravitational constant \cite{G}, and the fine structure constant \cite{Bouchendira}, for inertial sensing \cite{Canuel}, to test general relativity with part-per billion accuracy \cite{GravSME,redshift,GravAB}, and as a matter-wave clock \cite{CCC}, opening up new roads to testing the EEP. Such matter-wave interferometers use standing waves of laser light as diffraction gratings, leveraging the precision of laser wavelength measurement and avoiding use of material gratings, where antimatter atoms might annihilate. Unfortunately, they rely on Raman or Bragg transitions driven by nearly resonant lasers and are inefficient in using the available atom number. %This precludes operation with many atomic species.
Available continuous-wave \cite{Lymanalpha} or pulsed \cite{cool} lasers driving the Lyman-alpha line in hydrogen are not powerful enough unless collimated to a submillimeter beam radius. An atom interferometer using metastable hydrogen driven on the $2S\rightarrow 12P$ line with a 371\,nm laser has been demonstrated \cite{Heupel}. A similar scheme, using the $2S\rightarrow 3P$ line \cite{Fermilab} has been proposed, but bringing antihydrogen into the metastable state leads to further loss of scarce atoms. In this letter, we present a design that does not need a resonant laser. Instead, it uses a far-detuned, high-energy pulsed laser \cite{Gerlich}. The design also uses atom recycling to efficiently use the few available atoms. The interferometer can work with almost any atomic species, as well as electrons, protons, and their antiparticles.

\begin{figure}[t]
\centering
\epsfig{file=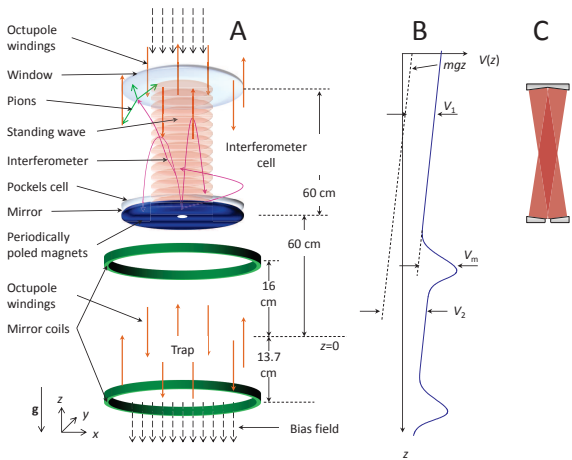,width=0.45\textwidth}
\caption{\label{interf} (A) Schematic. Atoms are extracted from the vertical magnetic trap (bottom) into the interferometer cell (top) by adiabatically lowering the trapping potentials, creating an antihydrogen fountain. The octupole is wound onto these walls of the vacuum chamber, which have an inner radius of 2.22\,cm. (B): Potential, not to scale. (C) Schematic of an off-axis multipass cell.
}
\end{figure}

The setup (Fig. \ref{interf} A) consists of two joined magnetic traps, the lower ``trap" region wherein antihydrogen atoms are produced and laser cooled, and the upper ``interferometer cell." These traps are similar to the one currently used by ALPHA \cite{AmoleNIM}, but oriented vertically. Atoms are laser-cooled to 20\,mK in the trap \cite{cool} and then adiabatically released into the interferometry cell. Interferometry is performed using a powerful off-resonant laser, retroreflected using a mirror that divides the interferometer cell and the trap. Atoms leaving the interferometer in the upwards-moving output leave the trap and annihilate at the top of the vacuum chamber. Spatially-resolved detection of annihilation products can count how many atoms leave the interferometer in the upper and lower output, respectively. This measures the phase shift between the interferometer arms and, thus, gravity.

Ramping down the trapping fields provides adiabatic cooling. A solenoid enclosing the entire setup (not shown) produces a homogenous, constant, vertical bias field $\textbf B_1$ of 1\,T. Octupole coils around the entire setup provide radial confinement by raising the field near the radial walls; mirror coils provide vertical confinement. A second solenoid surrounding only the trap region can be used to modify the bias field in the trap to $\textbf{B}_2$. Fig. \ref{interf} B shows the potential experienced by atoms on the axis. It consists of gravity $m g z$, where $m$ is the atom's mass and $z$ the vertical coordinate, a homogenous contribution $V_1$ by the overall solenoid that is modified to $V_2$ by the trap solenoid, and barriers of $V_m$ due to the mirror coils.

\begin{figure}
\centering
\epsfig{file=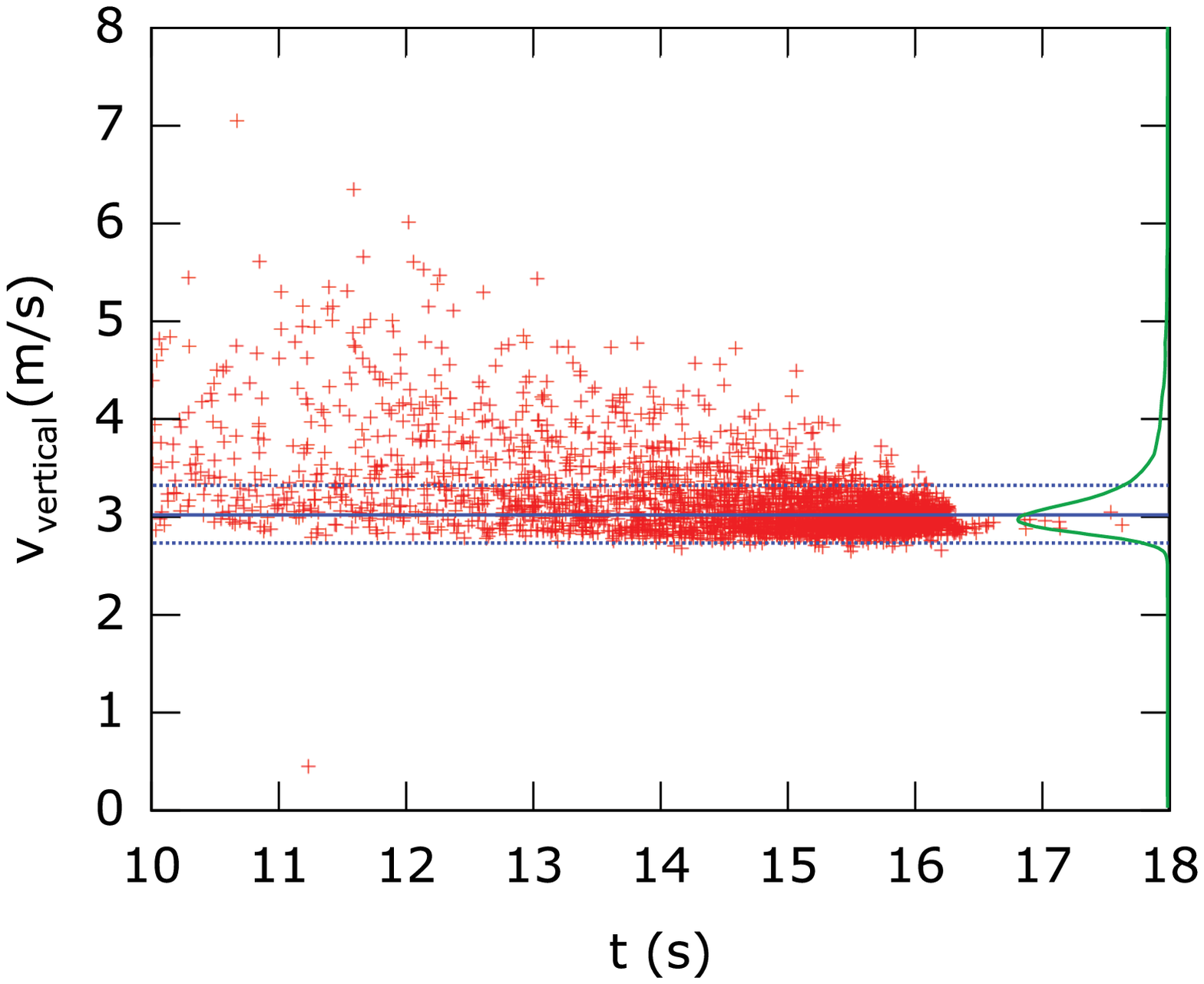,width=0.24\textwidth}
\epsfig{file=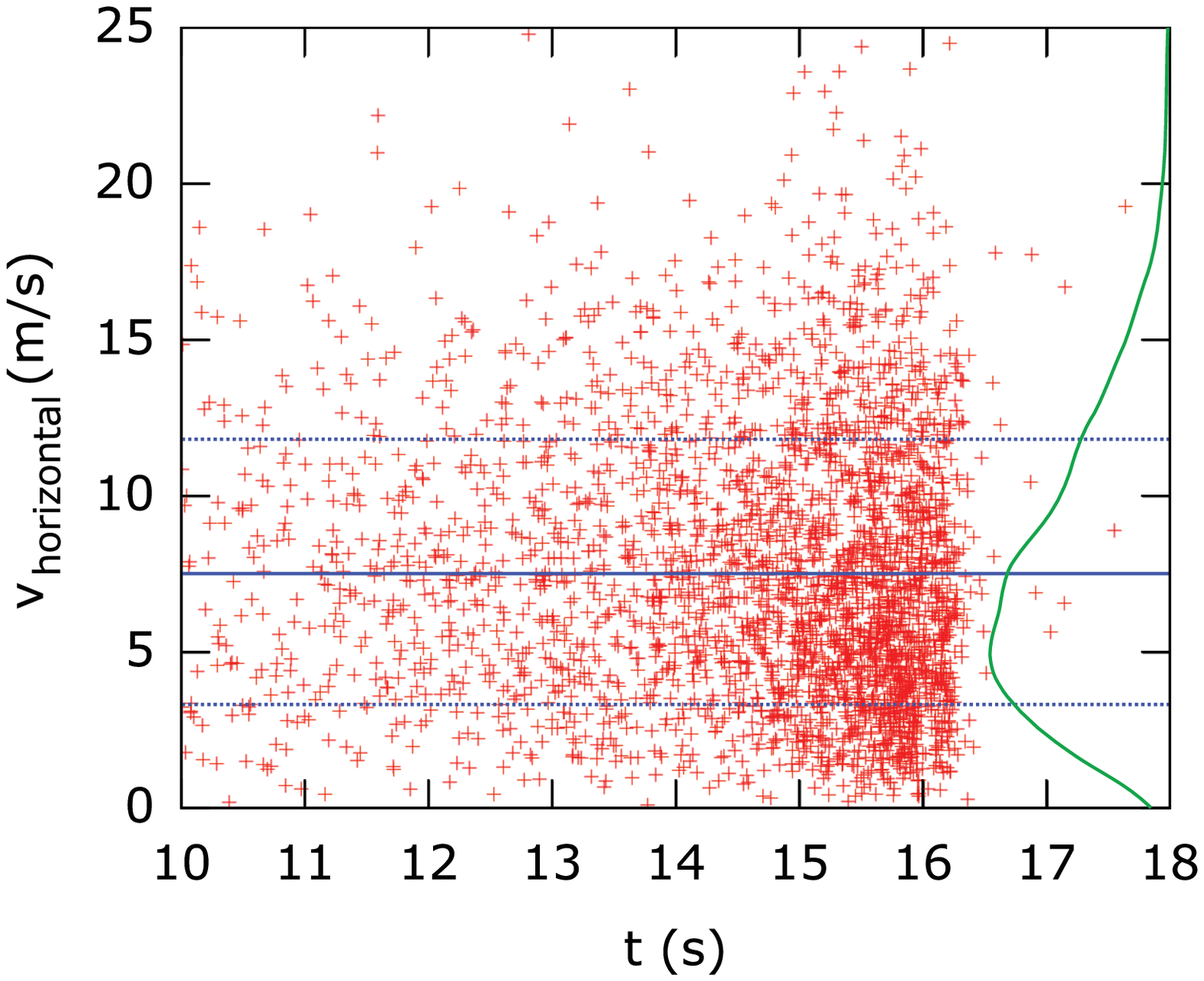,width=0.23\textwidth}
\caption{\label{solenoid} Vertical (left) and horizontal (right) velocity of extracted atoms, measured 40\,cm above the trap's center, versus time. Blue lines indicate the $1\sigma$ velocity spread, green lines the density of the velocity distribution. The fields are ramped exponentially with time constants of 40\,ms. For release, the upper mirror is ramped linearly within 16\,s %proportional to $-0.00096 + 0.0116(0.95 - s)$, where the time parameter $s$ goes from 0-1 during the 16\,s. The octupole is ramped
to 0.01 of its initial value, the lower mirror to 0.1, and the octupole to 0.15.}
\end{figure}

We use a pulsed Lyman-alpha laser for laser-cooling to a three-dimensional temperature of $\sim 20\,$mK, corresponding to a root mean square (rms) thermal velocity of $\sim 10$\,m/s \cite{cool}. During this time, the magnets are run at full fields, see \cite{AmoleNIM} for details on their design. In the second phase, that lasts 400\,ms, the octupole current is ramped down and the atoms are then allowed to expand to undergo adiabatic cooling. In a third phase, that lasts another 400\,ms, the lower and upper mirror coil currents are ramped down for further adiabatic cooling. After these phases, most antihydrogen atoms are still trapped. In the fourth phase, atoms are released over 16\,s. To achieve a nearly constant average vertical velocity, the trap solenoid is turned off completely while the upper mirror is ramped linearly. This results in particles entering the interferometer cell with the velocity distributions shown in Fig. \ref{solenoid}, with widths as narrow as 0.4 m/s rms vertically and 5 m/s horizontally. These figures can be improved further by optimizing the magnetic field configurations and ramp time constants. The interferometer cell is basically another magnetic trap. The overall potential seen by an atom depends on the radius coordinate $r$ as $\sqrt{V_6(r/\rho)^6+V_1^2}$, where $V_6$ and $\rho$ are constants.

The atoms enter the interferometer cell through an aperture. Without special precautions, a 1-cm aperture will pass most atoms. The area of the aperture can be reduced $q$ times if the trap potentials are ramped down $q$ times more slowly, without changing the velocity distribution. This follows from conservation of phase space density and is confirmed by our simulations. The atoms are prevented from colliding with the walls by periodically poled refrigerator magnets, see Fig. \ref{interf} A, which generate a repulsive potential that decays very fast with distance from the wall. Alternatively, we can use %a large central aperture with a slightly tilted laser beam or 
an off-axis multipass cell, see Fig. \ref{interf} C \cite{multipass}, which may also allow us to use a lower-powered laser.

The atom's fall under gravity and turn around $\sim 86$\,cm above the trap center before they reach the top of the interferometer cell, unless they are receiving an upwards momentum kick from the interaction with photons from the laser. Whenever the atoms reach the bottom, they are bounced back by the mirror coils with a probability of $P_b$, unless they disappear through the aperture and are then likely annihilated at the walls. The probability $P_b$ is controlled by the magnetic fields.

\begin{figure}
\centering
\epsfig{file=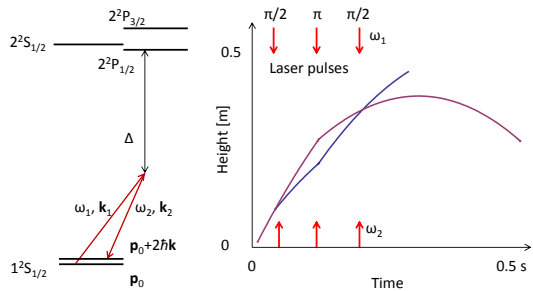,width=0.4\textwidth}
\caption{\label{bragg} Left: Bragg transition. Right: Space-time diagram of the Mach-Zehnder atom interferometer. A long pulse separation time $T=0.05$\,s has been chosen to clearly show the interferometer.}
\end{figure}

The atom interferometer is formed by the atoms' interaction with counterpropagating pulses from a laser whose wavelength is far off-resonant with any atomic transition, see Figure \ref{bragg} (left). Interaction with two laser beams transfers the atom from a state $|a,\textbf p\rangle$, where $a$ denotes the trapped $1^1S_{1/2}$ state of hydrogen and $\textbf p$ the atom's external momentum, into a state $|a,\textbf p+\hbar \textbf k_{\rm eff}\rangle$, where $\textbf k_{\rm eff}=\textbf k_1-\textbf k_2$ is the beams' effective wavevector. The Bragg condition, or energy and momentum conservation $
|\textbf p|^2/(2m)+\hbar \omega_1=|\textbf p+\hbar \textbf k_{\rm eff}|^2/(2m)+\hbar \omega_2,$ where $\omega_{1,2}$ are the laser frequencies, selects a certain initial momentum $\textbf p$ within a finite range given by the Fourier width of the laser pulses \cite{Bragg}. The interferometer sequence is repeated at a rate of, e.g., 20\,Hz. %while the atom is accelerating under gravity. Thus, every atom will encounter the laser beam while meeting the Bragg condition.
The two counterpropagating beams are generated by retroreflection on a mirror (Fig. \ref{interf}) with two passes through a Pockels cell. Ramping the phase shift introduced by the cell controls $\om_1-\om_2$. This has the advantage that no laser beams need to pass the trap region, allowing greater flexibility in the placement of components there. If the trap offers uninhibited optical access from both sides, however, we may avoid the use of optical elements inside the vacuum chamber.

If the Bragg condition is satisfied, the probability of the Bragg transition is given by $P_{ab} = \sin^2(\Phi_R/2)$, where $\Phi_R=\int \Om^{(2)}dt$ is given by the two-photon Rabi frequency $\Om^{(2)}$. A $\Phi_R=\pi/2$-pulse creates an equal superposition of wave packets that separate vertically with a recoil velocity of $\hbar k_{\rm eff}/m$; a $\Phi_R=\pi$-pulse acts as a mirror. For a far-detuned infrared laser, $\Om^{(2)}=\alpha I/(2\eps_0\hbar c)$ is given by the atom's dc polarizability $\alpha$, the laser intensity $I$ and the vacuum permittivity $\eps_0$. For hydrogen, $\alpha=(9/2) 4\pi \eps_0 a_0^3$ exactly, so that $\Om^{(2)}=9\pi a_0^3 I/(\hbar c)$, where $a_0$ is the Bohr radius. Since the dc polarizability is nonzero for any atom, the interferometer can work with any species.

\begin{figure}
\centering
\epsfig{file=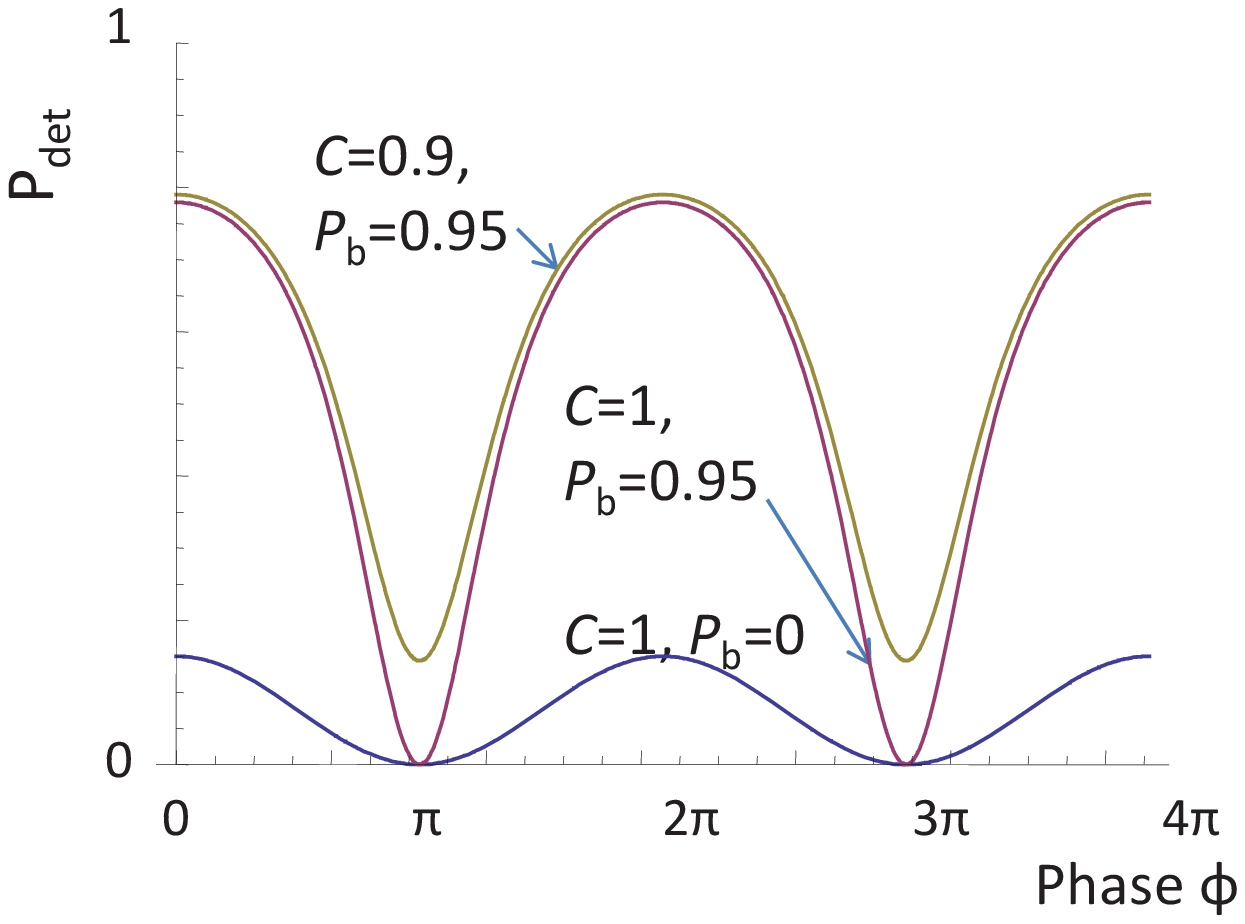,width=0.23\textwidth}
\epsfig{file=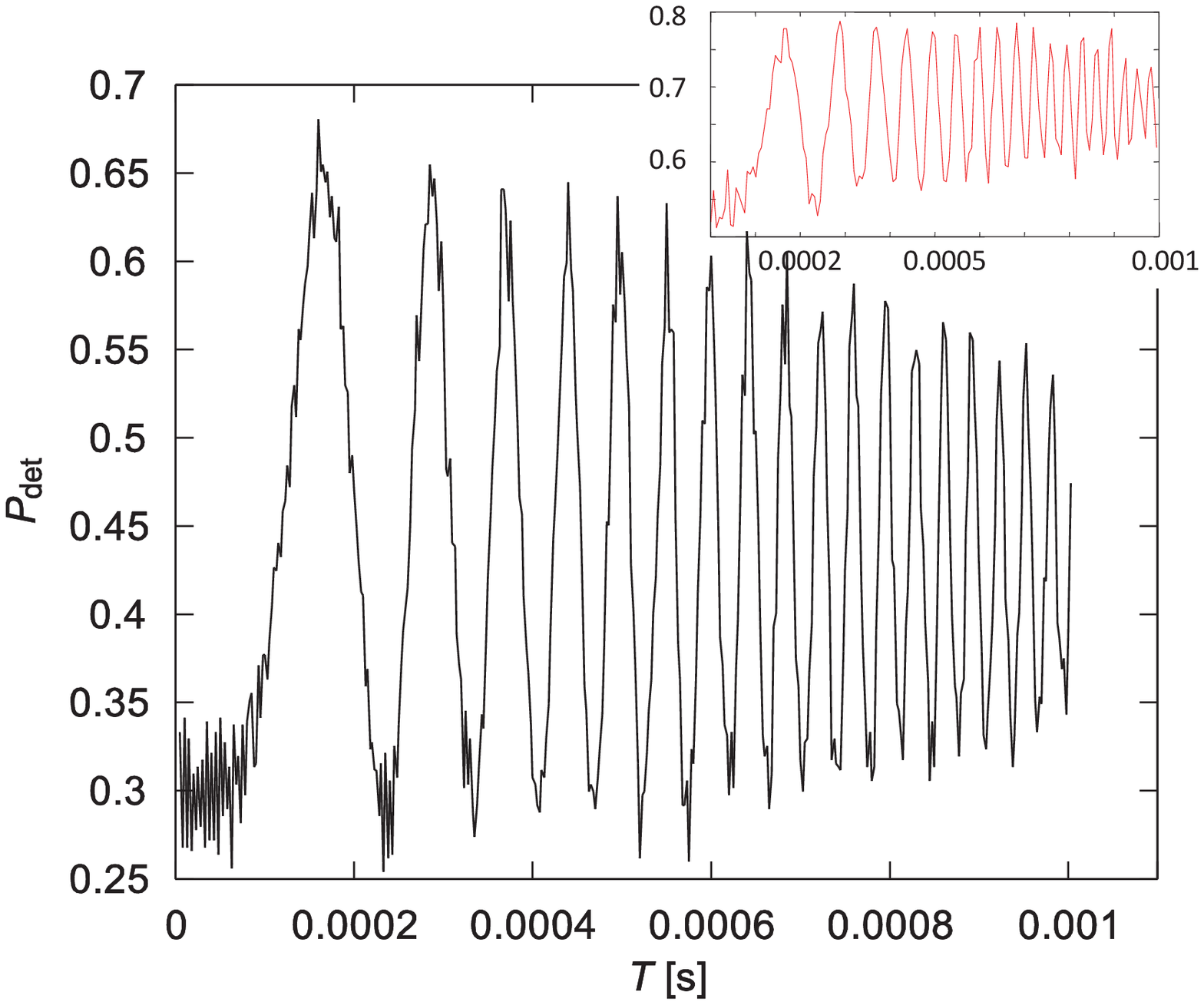,width=0.23\textwidth}
\caption{\label{Recycling} Left: Fringes of a simple interferometer with $A=15\%$, roughly what can be achieved with a 10-mm radius laser beam in a 25-mm radius trap. Atom recycling leads to higher visibility and sharpens the features. Right: Simulation of the full atom interferometer. The number of atoms detected at the top of the interferometer cell versus pulse separation time $T$ shows the expected $\sin^2(kgT^2)$ signature. Inset: simulation taking into account a 5-mm diameter aperture in the mirror, with 256\,s adiabatic release time.}
\end{figure}

A combination of $\pi/2-\pi-\pi/2$ pulses, spaced by intervals $T$, split and recombine the matter waves so that they interfere, Fig. \ref{bragg} (right). The probability $P_\uparrow$ of detecting the atom at, {\em e.g.}, the upper output of the interferometer are given by the phase difference $\phi$ accumulated between the matter waves on the two paths \cite{AI},
\begin{equation}\label{phi}
\phi= (\textbf k_{\rm eff} \cdot \textbf g) T^2.
\end{equation}
To leading order, this is independent of the atom's initial velocity and position. Detecting the atoms in the upper and lower output of the interferometer measures the phase difference and thus $\textbf g$. The population in the upper output can be written as $P_\uparrow = A \cos^2(\phi/2) +B$. An ideal interferometer would have a contrast $C=A/(A+2B)$ of one. In practice, this ideal contrast is not realized, e.g., when laser pulses miss the atom. In our proposal, however, such atoms keep orbiting in the trap and thus have a chance of $P_b$ to encounter the laser beam again and take part in an interferometer. In a simple model, the total probability that an atom is eventually scattered upwards is given by a geometric series
\begin{equation}\label{Pup}
P_{\rm det}=P_\uparrow \sum_{n=0}^\infty (1-P_\uparrow)^nP_b^n.
\end{equation}
Such atoms reach the top of the interferometer cell, where they annihilate with the walls and are thus detected. %Position-sensitive detection of pions helps to disciminate against background counts.
Figure \ref{Recycling} (left) shows that fringes of near-unity peak-to-peak amplitude are obtained. %even when the one for a single atom interferometer is, e.g., just $A=0.15$. %This makes it possible to operate with dilute antihydrogen samples. Fringes even steeper than those of a perfect conventional interferometer can be reached.
Due to the increased slope, the interferometer can in principle surpass the sensitivity limits of a single interferometer for a given atom number.

We simulate the interferometer for the dimensions shown in Fig. \ref{interf}. The simulation fully takes into account the geometry of the trap, the laser beam, and all magnetic fields; the 3-dimensional motion of the atoms, and the quantum mechanics of the beam splitters. It starts with tracing the paths of a laser-cooled sample of antihydrogen at 20\,mK in the trap for 0.1\,s and then simulating the adiabatic release from the trap (Fig. \ref{solenoid}). The laser beam has 1064-nm wavelength and 1\,cm radius with a flat-top intensity profile. The pulses have a Gaussian time envelope with a $\sigma=250\,$ns time constant and a $\pi-$pulse energy of 7.4\,J \cite{multiphoton}. %They can be obtained by amplifiying a narrow-linewidth continuous-wave laser using commercial pulsed, diode-pumped Neodymium-doped Yttrium-Aluminum Garnet (Nd:YAG) laser amplifiers \cite{NGO}.
The atom-light interaction is modeled by numerically integrating the Schr\"odinger equation using the $|a,2n \hbar k\rangle$ ($n=-5,\ldots 5$) states as basis states, fully accounting for the Doppler shift of the laser frequencies as seen by the moving atoms.  %We assume that the quantum coherence is lost between two interferometer sequences. Between light pulses, the wave packets' evolution under gravity and the potentials created by the magnets is modeled using the Wentzel-Kramers-Brillouin method.

Fig. 4 (right) shows the interference fringes obtained. Fringes show a high contrast of $\sim 35\%$. The contrast decay with pulse separation time $T$ is relatively mild, since the trapping potentials confine about 80\% of all atoms to within the area illuminated by the laser beam. The observed contrast decay is due to magnetic field gradients caused by the mirror coils. It can be reduced by using higher-multipole mirror coils whose field decays faster with distance. A laser of shorter wavelength, e.g., 532\,nm, will increase the initial contrast to $\sim 50\%$, as the larger recoil velocity has a more favorable ratio to the vertical velocity spread. Short wavelengths also lead to a larger measured signal, Eq. (\ref{phi}), allowing better resolution. If an off-axis multipass cell or a tilted beam is not used, the absence of a reflected beam over the aperture's area will lead to a slight loss of contrast, see the inset in Fig. \ref{Recycling}.

The scarcity of atoms means that the performance of the experiment is likely to be limited by noise. Currently, approximately four atoms can be trapped and detected per hour, without laser cooling, with at least 1000\,s confinement time.
With the upgraded antiproton ring ELENA, this rate can theoretically be sustained continuously to make $\sim 3\times 10^3$\,atoms/month. Allowing for equipment downtime and losses during laser cooling, we here assume 250 detected atoms available per month.

As a ``basic" scenario, we discuss an interferometer with a pulse separation time of $T=1\,$ms. The shot-noise limit with 250 detected atoms and a contrast of $>50\%$ leads to a statistical resolution of $\delta g/g=0.8$ parts per thousand (ppt), see Tab. \ref{noise}. Vibration of the retroreflection mirror typically amounts to accelerations of $\sim 10^{-3}\,$m/(s$^2\sqrt{\rm Hz}$) at the kHz-frequencies $\om/(2\pi)$ around $1/T$ in a lab \cite{Hensley} but may be ten times as high in the antihydrogen trap. Their effect is calculated by convoluting the spectrum of vibrational motion with the $\sin^2(\omega T/2)/(\omega T)$ sensitivity function of the interferometer. ALPHA uses localized detection with a silicon detector to suppress background counts to a level of  $1.7\times 10^{-3}$ per second \cite{AmoleNature,AndresenNatPhys} or about 0.01 within the 6\,s during which most anti-atoms are released from the trap (Fig. \ref{solenoid}). Noise due to background counts is thus $\sqrt{0.01}$ as large as atom shot noise. Mirror coil field gradients %on axis are given by $B_0 R^3/(z^2+R^2)^{3/2}$ with $B_0\simeq 0.01\,$T and $R\simeq 5\,$cm. If the interferometer takes place at $z = (50\pm 5)$\,cm above the coil, this field
produce offsets of errors of $\delta g/g=23$\,ppt that can be calibrated, and a 9\,ppt fluctuation per atom, as the height at which the atom encounters the laser beam is random. The bias field $B_1$ has gradients in practice \cite{magnet}. For $B_1=0.05\,$T and a typical scale of field fluctuations of 10\,cm, fluctuations need to be less than $2\times 10^{-5}B_1$ to reduce the noise from random atom positions. %Noise due to the laser system, {\em e.g.}, laser intensity fluctuations, Pockels cell phase noise, and other effects, is probably negligible. We leave this for future study.

\begin{table}
\caption{\label{noise} Noise sources for the basic and advanced scenario, times month$^{1/2}$.}
\begin{tabular}{ccc}\hline\hline
Source & basic [ppt] & advanced [ppm] \\ \hline
Atom shot noise & 0.8 & 0.7 \\
Atom motion in mirror field & 0.6 & 1 \\
Atom motion in solenoid field & 0.4 & 0.7 \\
Vibrations & 0.4 & 0.1 \\
Background counts & 0.1 & 0.1 \\\hline
Total & 1.2 & 1.4 \\ \hline \hline
\end{tabular}
\end{table}

For an ``advanced" scenario, we assume $10^3$ atoms/month, a laser wavelength of $532\,$nm, a pulse separation time of 20\,ms, and that magnetic fields can be flattened 300-fold. This major challenge may be met with, {\em e.g.}, additional coils, lowering the overall fields, or calibration. Vibrational noise is reduced by interferometric read-out of the mirror vibration relative to a quiet reference mirror outside the trap, which will lead to less than $\sim 10^{-4}\,$m/(s$^2\sqrt{\rm Hz}$) acceleration at the relevant frequencies \cite{Hensley}. Everything else being equal, we obtain the estimates in Tab. \ref{noise}.

%We note that systematic effects that limit part-per billion measurements such as Coriolis force and rotations, laser wavefront curvature, Gouy phase, and laser wavelength stability are negligible here \cite{Peters,Bouchendira,CCC}. We assume that the interferometer has been calibrated with hydrogen atoms. This should take out permanent errors, assuming CPT symmetry holds for the atoms' response to the lasers and trapping fields.

We have presented a project to verify the EEP for antihydrogen. %, initially to $10^{-2}$, later to $2\times 10^{-6}$ resolution.
%Its design meets the challenges presented by the scarcity of atoms, the nonlocalized atom source, high thermal velocity of the atoms, and the Lyman-alpha wavelength of 121\,nm. The far off-resonant Bragg beam splitter will work with any species of atom, separately or simultaneously.
Using the same laser and trap geometry, the experiment can work with any atoms, but also electrons, protons as well as their antiparticles using the Kapitza-Dirac effect for light-pulse beam splitters \cite{Batelaan}. These advances will allow building an electron/positron interferometer and matter-wave clock, which can be used to verify the EEP for charged particles as a null redshift experiment, and for a precision measurement of the particle's masses \cite{CCC}. Charged particles would be trapped in a weak Penning trap; fields due to patch charges on the surface could be suppressed by free electrons on a thin helium film \cite{Hefilm}.

We thank W. Bertsche, E. Butler, M. Fujiwara, J. Hangst, P. Haslinger, N. Madsen, and T. Tharp for detailed discussions, and D. Kaplan for stimulating discussions in the initial phase of this work. This work has been supported by the David and Lucile Packard foundation, the Department of Energy, the National Aeronautics and Space Administration, and the National Science Foundation.


\begin{thebibliography}{99}
\bibitem{multiphoton} Under these conditions, the probability of multiphoton ionization is $<10^{-61}$ per pulse \cite{Bebb}.
\end{thebibliography}
\end{document}